\documentclass[aps,preprint,showpacs,floatfix,superscriptaddress]{revtex4}
\usepackage{graphicx}

\begin{document}
\flushbottom

\title{Local superconducting density of states of ErNi$_2$B$_{2}$C}
\author{M. Crespo}
\affiliation{Laboratorio de Bajas Temperaturas, Departamento de
F\'isica de la Materia Condensada \\ Instituto de Ciencia de
Materiales Nicol\'as Cabrera, Facultad de Ciencias \\ Universidad
Aut\'onoma de Madrid, 28049 Madrid, Spain}
\author{H. Suderow}
\affiliation{Laboratorio de Bajas Temperaturas, Departamento de
F\'isica de la Materia Condensada \\ Instituto de Ciencia de
Materiales Nicol\'as Cabrera, Facultad de Ciencias \\ Universidad
Aut\'onoma de Madrid, 28049 Madrid, Spain}
\author{S. Vieira}
\affiliation{Laboratorio de Bajas Temperaturas, Departamento de
F\'isica de la Materia Condensada \\ Instituto de Ciencia de
Materiales Nicol\'as Cabrera, Facultad de Ciencias \\ Universidad
Aut\'onoma de Madrid, 28049 Madrid, Spain}
\author{S. Bud'ko}
\affiliation{Ames Laboratory and Departament of Physics and Astronomy \\
Iowa State University, Ames, Iowa 50011, USA}
\author{P.C. Canfield}
\affiliation{Ames Laboratory and Departament of Physics and Astronomy \\
Iowa State University, Ames, Iowa 50011, USA}

\begin{abstract}
We present local tunnelling microscopy and spectroscopy
measurements at low temperatures in single crystalline samples of
the magnetic superconductor ErNi$_2$B$_2$C. The electronic local
density of states shows a striking departure from s-wave BCS
theory with a finite value at the Fermi level, which amounts to
half of the normal phase density of states.
\end{abstract}

 \date{\today}
\maketitle

During the last decade, many studies of the quaternary rare earth,
nickel borocarbides have provided important insight into the
intriguing and long standing problem of coexistence of magnetic
and superconducting orders \cite{Bulaevski85,Mueller01}. In
particular, compounds with RNi$_2$B$_2$C and R = Y, Lu, Tm, Er, Ho
and Dy, are superconducting although their magnetic properties are
very different. The compounds with R = Y and Lu bring only the
superconducting character into play and have the highest critical
temperatures of the series (T$_c$ = 15.5 and 16.5 K respectively).
Those with R = Tm, Er, Ho or Dy are magnetic superconductors with
smaller but still sizable critical temperatures (T$_c$ = 11, 10.5,
8.6 and 6 K respectively) \cite{Mueller01}. In all of them, high
quality single crystalline samples have been grown in several
laboratories, and there are in addition experiments and
calculations of the electronic bandstructure
\cite{Mueller01,Dugdale99,Rhee95,Divis00}. All these circumstances
have promoted a notable interest in this family, considered by
many as a "toy-box" where an impressive range of new physical
effects are to be found \cite{Canfield98}.

The compounds ErNi$_2$B$_2$C and TmNi$_2$B$_2$C have very similar
T$_c$'s, which are the highest among the magnetic compounds of the
series. Nevertheless, their magnetic properties are radically
different. TmNi$_2$B$_2$C is paramagnetic down to 1.5~K, where it
transits to an antiferromagnetic state with an incommensurate
wavevector \cite{Movshovich94,Lynn97}. The ordered magnetic moment
amounts to 3.8 $\mu_B$, and is much smaller than the one of
ErNi$_2$B$_2$C (7.2 $\mu_B$), which transits already below T$_{N}$
= 6~K, to an also incommensurate spin density wave, and below
T$_{WF}$ = 2.3~K to a peculiar weak ferromagnetic state where one
of each twenty Er spins are aligned resulting in a small net
magnetic moment of 0.33$\mu_B$ per Er
atom~\cite{Zaretsky95,Canfield96,Choi01,Jensen02,Kawano02,Walker03,Jensen04}.
Several remarkable new effects have been found to occur below
T$_{WF}$ as a consequence of weak ferromagnetism. For instance,
the vortex lattice tilting away by a small but measurable amount
(up to 1.6 degrees) from a magnetic field applied perpendicular to
the magnetic moments, i.e. in the plane of the tetragonal crystal
structure~\cite{Yaron96}, or an increase of the critical
current\cite{Gammel00}. There are however no available data of the
electronic excitation spectrum. Here we present measurements of
the superconducting electronic density of states using local
tunnelling spectroscopy. We find that ErNi$_2$B$_2$C exhibits yet
another interesting property, absent in other nickel borocarbide
compounds, namely an anomalously high amount of low lying
electronic excitations in the local density of states.

\begin{figure}[ht]
\includegraphics[width=7cm,clip]{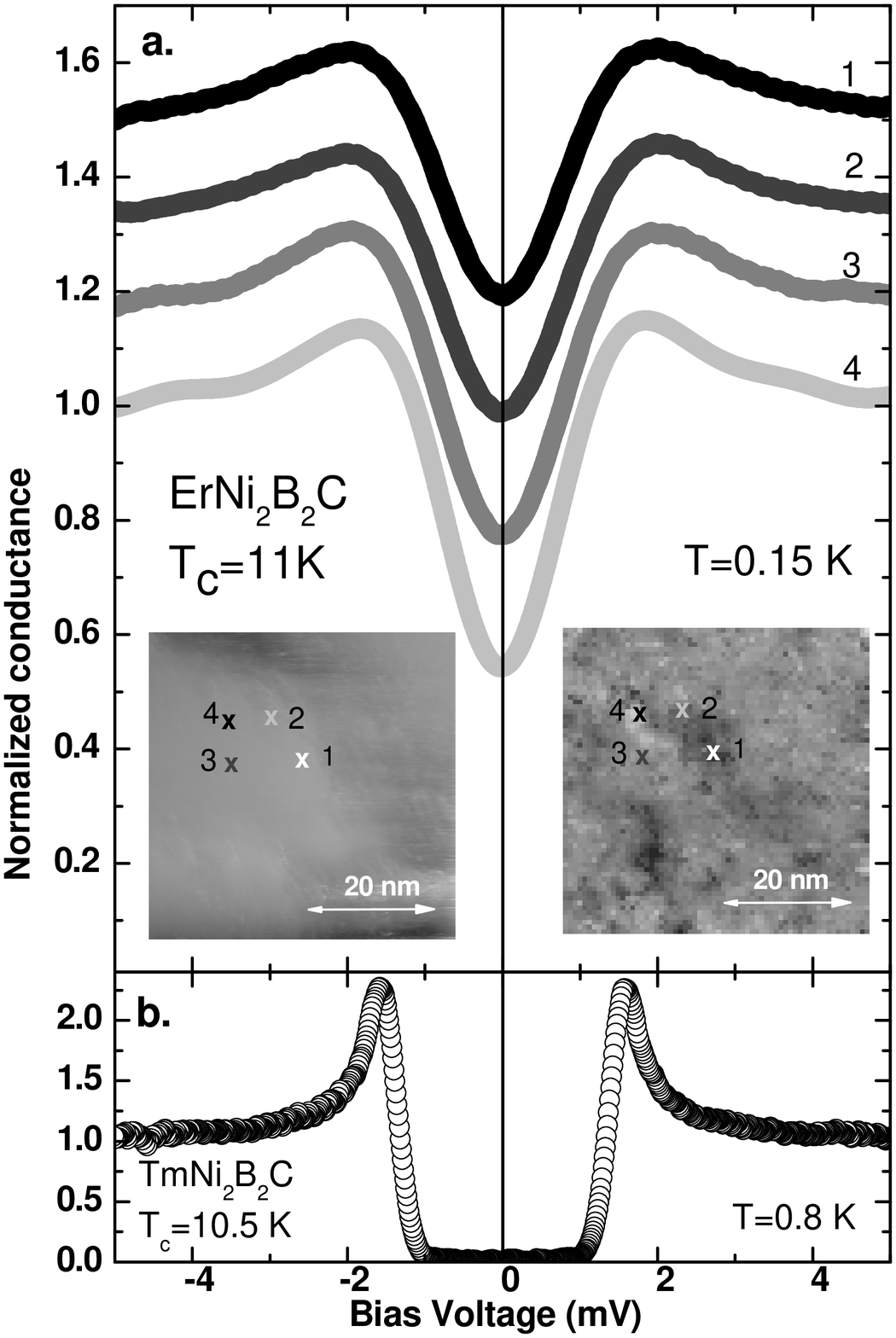}
\vskip -0.2cm \caption{In a. we show several characteristic
tunnelling spectra, taken at 0.15~K, and shifted by +0.18
(tunnelling resistance R$_T$ = 10 M$\Omega$). In the insets we
show typical topography (left) and STS (right) images found in
ErNi$_2$B$_2$C. The full range in contrast from black to white
represents, respectively, a height difference of 20 nm in
topography (left) and a change in the zero bias conductance of
20\% in STS (right). Crosses and numbers give points where the
tunnelling spectra shown in the figure (top) were obtained. For
comparison we show in b. data of the similar material
TmNi$_2$B$_2$C, published in \protect\cite{Suderow01}, and taken
at 0.8 K.} \label{fig1}
\end{figure}

We use a very similar STM set-up as in previous measurements in
other superconducting nickel borocarbide materials
(TmNi$_2$B$_2$C\cite{Suderow01}, and the non-magnetic
YNi$_2$B$_2$C, LuNi$_2$B$_2$C with T$_c$= 15.5~K and 16.5~K
\cite{Martinez03b}). The spectral resolution of the experimental
set-up has been improved through the measurement of low critical
temperature superconductors as Al and PrOs$_4$Sb$_{12}$ (T$_c$ =
1.12 K and 1.85 K respectively), where we were able to obtain
clean spectra with a negligible conductance at zero bias and a
resolution in energy of 15 $\mu$eV \cite{Rodrigo04,Suderow04}. In
addition, we have added a blade to the sample holder, so that the
samples can be broken in-situ, at low temperatures under cryogenic
ultra high vacuum. We mount small (about 1x1x5 mm$^3$) single
crystalline samples into the sample holder. After cooling, we
break the samples with the blade and approach a clean gold tip
mounted on a piezotube with a scanning range of 600x600 nm$^2$. An
x-y table permits macroscopic positioning of the tip over the
sample and allows to study scanning windows in macroscopically
different regions of each sample, without heating the whole
set-up. As demonstrated in all other previous STM studies made in
the borocarbides by us and other
groups\cite{Martinez03b,Suderow01,Sakata00,deWilde97}, the samples
show a conchoidal fracture, without a clear cleaving plane.
Accordingly, the surface of these samples, as measured with the
STM, has large relatively flat regions, showing at the smallest
scales a small but finite roughness
\cite{Martinez03b,Suderow01,Sakata00,deWilde97}. Therefore,
tunnelling is typically made at arbitrary directions, so that it
becomes especially important to study a large number of samples in
macroscopically different regions. We successfully broke in-situ
fourteen samples, verifying after the measurements the good
quality of the surfaces with a SEM, and the tunnelling plane using
X-ray scattering. Surfaces in and out of plane, as well as in
intermediate directions have been probed. In each sample, we
studied between five and ten different scanning windows. We took
current-voltage curves in different positions, and also studied
systematically scanning tunnelling spectroscopy images (STS),
obtained by making a current-voltage curve at each point of a
64x64 array, and subtracting the conductance well within the
quasiparticle peaks from the conductance at high bias (see for
more details
Refs.~\cite{Martinez03b,Suderow01,Sakata00,deWilde97}). In all
cases, we found surfaces of high quality, as in previous work, but
with dramatically different tunnelling spectra for the case of
ErNi$_2$B$_2$C, as will now be discussed.

At the lowest temperatures, tunnelling spectra as shown in
Fig.~\ref{fig1}a are always found over large areas of the surface.
These curves differ radically from the predictions of s-wave BCS
theory. Instead of a wide energy range with a zero conductance, we
find a finite conductance at zero and low bias, which corresponds
to about one half of the high bias conductance. We have included
in Fig.~\ref{fig1}b the spectra obtained in the very similar
superconductor TmNi$_2$B$_2$C, published in Ref.\cite{Suderow01}.
In TmNi$_2$B$_2$C, the form of the spectra follow the BCS, s-wave
theory quite closely with a negligible conductance at the Fermi
level. In the non-magnetic compounds YNi$_2$B$_2$C and
LuNi$_2$B$_2$C, we also always find a negligible conductance at
the Fermi level~\cite{Martinez03b}.

In Fig.~\ref{fig1}a we show a typical STS image (bottom right),
which represents the changes found in the zero bias conductance in
the 400~x~400~nm$^2$ window whose topography is represented in the
bottom left image of Fig.~\ref{fig1}a. Clearly, the spectra remain
with the same form in the whole scanning range, apart small
variations of 10-20\% (grey scale in bottom right image of
Fig.~\ref{fig1}a), which are not correlated to any patterns in the
topography. Similar images and spectra have been found in all
fourteen in-situ broken single crystalline samples.

At temperatures much below T$_c$, the local tunnelling conductance
is a direct measure of the local density of states N$_{loc}$(E).
Its form depends on the superconducting properties of the surface,
its detailed electronic structure, and on the tunnelling plane.
Our experiment evidences a high density of states at the Fermi
level together with a well defined V-shaped increase between zero
bias and the peak observed at $\Delta$ = 1.8 meV
(Fig.~\ref{fig1}a). This value is similar, although slightly
larger, to the gap expected from most simple BCS theory
($\Delta_0$~=~1.73k$_B$T$_c$), which amounts to 1.65~meV, and to
the value found in TmNi$_2$B$_2$C (1.45~meV) \cite{Suderow01}.

\begin{figure}[ht]
\includegraphics[width=7cm,clip]{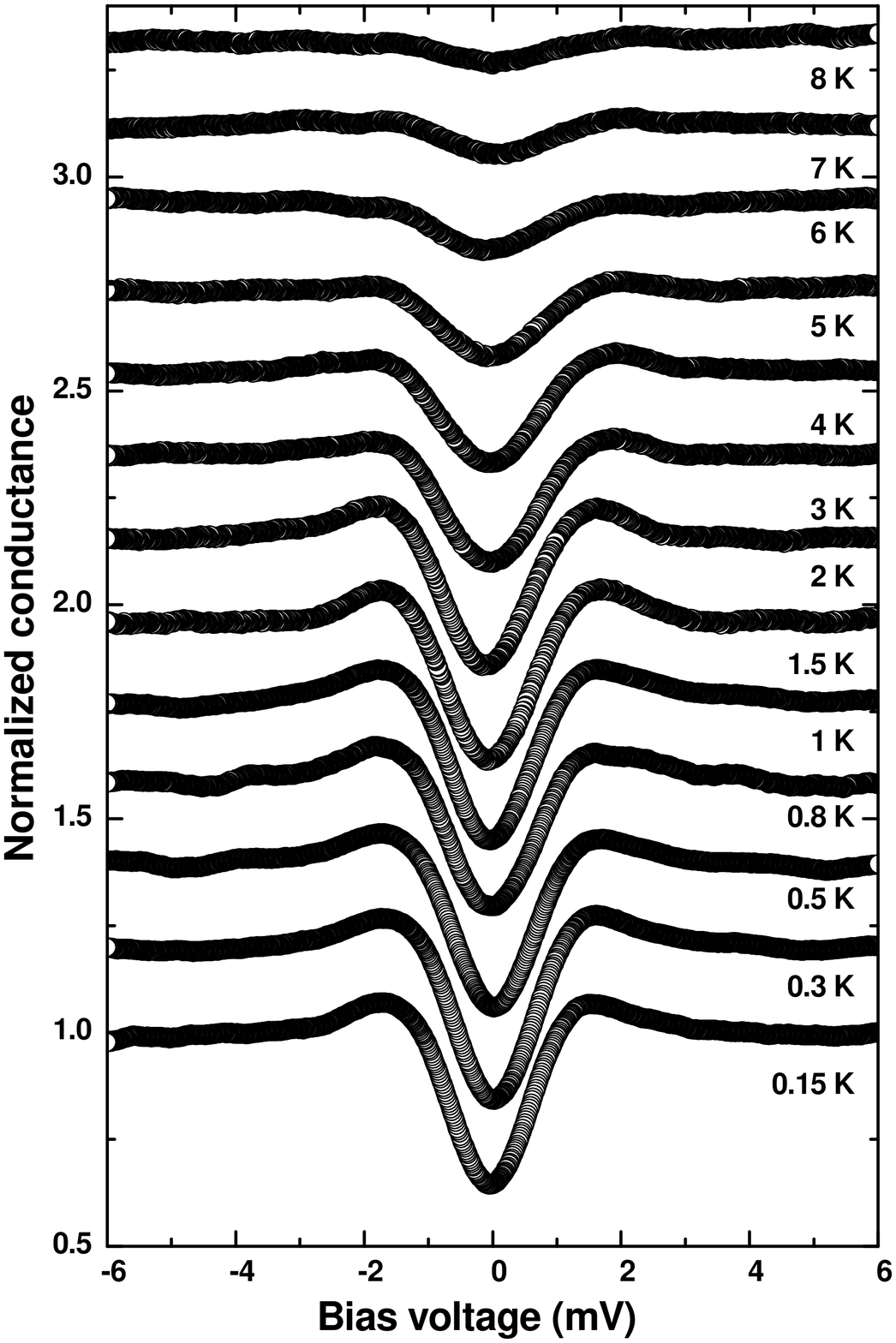}
\vskip -0.5cm \caption{The figure shows a representative
temperature scan. The spectra maintain its form in the whole
temperature range, without showing important changes at the
magnetic transitions (T$_{N}$ = 6~K and T$_{WF}$ = 2.3~K).}
\label{fig2}
\end{figure}

\begin{figure}[ht]
\includegraphics[width=7cm,clip]{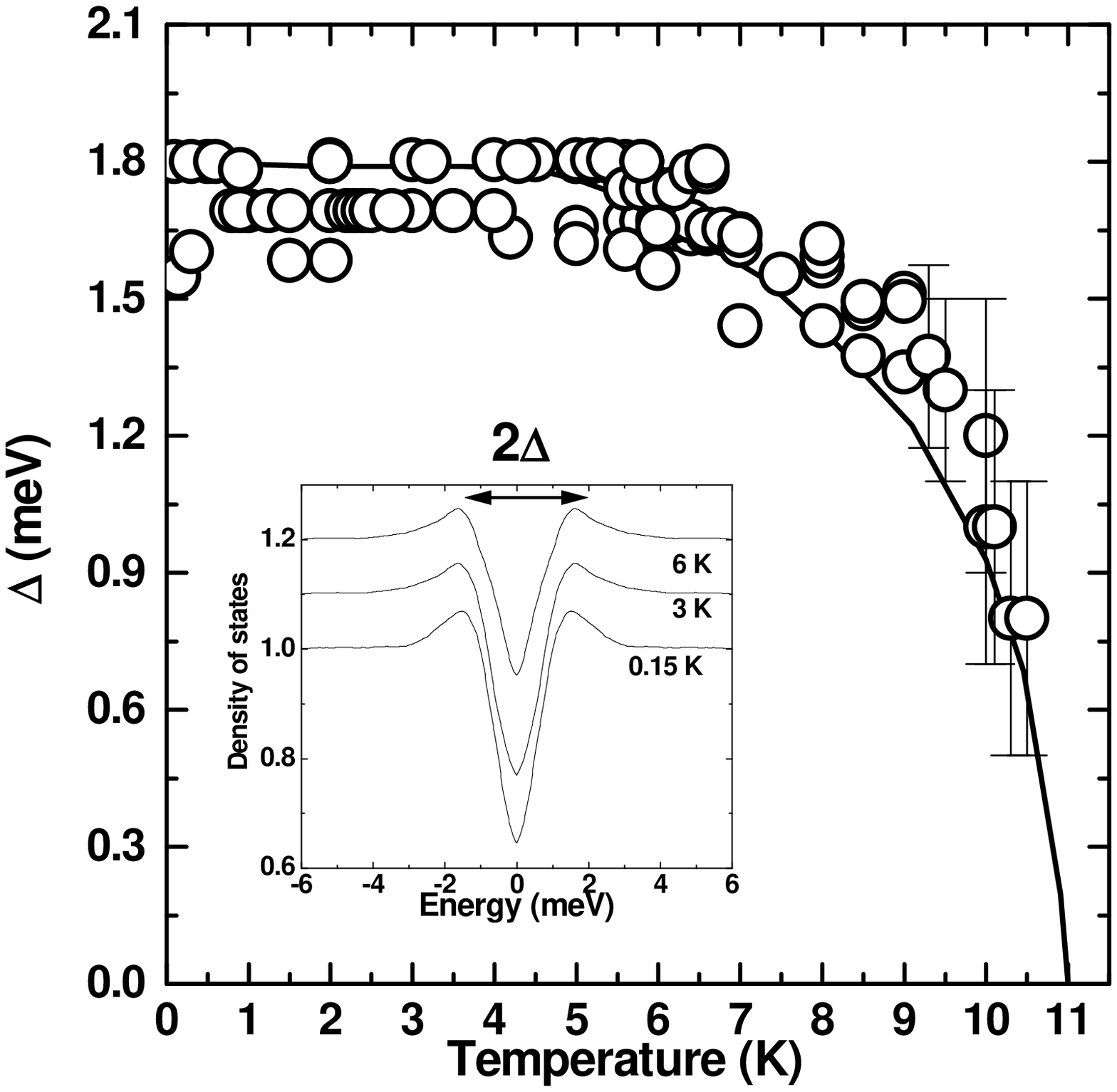}
\vskip -4cm \caption{The position in energy of the quasiparticle
peaks in the density of states, $\Delta$, is plotted as a function
of temperature. Solid line is the temperature variation of the
superconducting energy gap within BCS theory. In the inset we show
an example of the superconducting density of states at three
different temperatures, obtained by de-convoluting the density of
states from tunnelling conductance spectra. $\Delta$ is obtained
as schematically shown by the arrows.} \label{fig3}
\end{figure}

The temperature dependence of the superconducting spectra has been
also followed making temperature scans in each sample in different
positions. The changes in the tunnelling conductance when
increasing temperature are shown in a representative scan in
Fig.~\ref{fig2}. The form of the local density of states is simply
smeared out by temperature. The position of the peak in the local
density of states $\Delta$ can be followed as a function of
temperature by de-convoluting the density of states from the
tunnelling conductance. It is interesting to note that, as shown
in Fig.\ \ref{fig3}, the temperature variation of $\Delta$ follows
well BCS prediction for the superconducting gap.

Therefore, our result evidences the presence of an important
amount of ungapped excitations on the surface, which shows no
sample nor orientational dependence within our experimental
resolution, and of gapped excitations. The latter have a
distribution of values of the superconducting gap, which produce a
V-shaped density of states, and have a maximum value, $\Delta$
that decreases with temperature following BCS theory.

Clearly, the onset of magnetic order in ErNi$_2$B$_2$C at T$_N$ =
6 K and T$_{WF}$ = 2.3 K does not greatly change the local
tunnelling conductance. It is interesting to compare to
TmNi$_2$B$_2$C, where no changes are found in the tunnelling
spectra when crossing T$_N$ = 1.5 K. In the antiferromagnetic
phase, the local magnetic field indeed averages to zero in
distances much shorter than the superconducting coherence length,
as $\xi_0$ = 12 nm \cite{Cho95}, whereas the magnetic moment
changes sign approximately each 2 nm \cite{Eskildsen98}. As a
consequence, the superconducting density of states does not appear
to be influenced by the onset of magnetic order and follows well
expectations from simple BCS theory \cite{Suderow01}. In
ErNi$_2$B$_2$C, the superconducting coherence length is of the
same magnitude ($\xi_0$ = 13.5 nm), and the antiferromagnetic
modulation of the magnetic moment occurs at an even smaller length
scale, within the unit cell \cite{Zaretsky95,Canfield96}. So that
we should not expect a strong change of the superconducting
density of states of ErNi$_2$B$_2$C at T$_N$ due to the local
magnetic field. Nevertheless, we should point out that T$_N$ is
rather high in ErNi$_2$B$_2$C, well above T$_c$/2, so that the
temperature induced smearing in the tunnelling conductance is
significant and the determination of the local density of states
is not as precise as at lower temperatures. In our experiment, at
6 K, we cannot resolve the density of states within the
quasiparticle peaks better than about 30\%. An effect which may
produce only small changes in the local density of states and fall
within this error bar is the small gap which certainly opens at
the Fermi level at T$_N$ in ErNi$_2$B$_2$C, because the magnetic
ordering wavevector of ErNi$_2$B$_2$C nests a small part of the
Fermi surface \cite{Dugdale99,Zaretsky95}. As regards the peculiar
weak ferromagnetic order of ErNi$_2$B$_2$C, it is expected to
create a finite magnetic field below T$_{WF}$ = 2.3 K. Estimates
give values which are rather small, of the same order or slightly
larger than H$_{c1}$ (between 0.05 T and 0.1 T
\cite{Canfield96,Ng97,Gammel00,Chia04}) and cannot be expected to
produce changes greater than 10\% on the amount of low energy
excitations through magnetic pair breaking \cite{Suderow02}. On
the other hand, the possible appearence of an intriguing vortex
lattice at zero field due to this small magnetic field has been
discussed by several authors \cite{Ng97,Chia04}. Clearly, it does
not show up on surface scanning tunnelling spectroscopy, within
the scattering of the data shown in Fig.\ \ref{fig1}.

Note that in all cases, we did follow the superconducting spectra
up to the bulk T$_c$. This rules out simple impurity pair breaking
effects as the origin of the low energy excitations. In principle,
pair breaking due to impurities or defects is especially
significant in antiferromagnetic superconductors, where Anderson's
theorem is violated and even non-magnetic impurities act as pair
breakers \cite{Nass82,Bulaevski85}. However, T$_c$ is greater than
T$_N$ in ErNi$_2$B$_2$C. Moreover, pair breaking is always
associated to a decrease of T$_c$ related to the amount of
excitations created through pair breaking within the gap, which
increases the zero bias conductance \cite{Nass82,Bulaevski85}.
Therefore, this can be fully ruled out in ErNi$_2$B$_2$C where the
form of the tunnelling conductance is not sample dependent and,
what is more important, survives up to the bulk T$_c$ in all
measured samples.

On the other hand, the suppression of T$_c$ by intrinisic magnetic
pair breaking due to the local moment of the rare earth has been
largely discussed to explain the overall behavior of T$_c$ along
the RNi$_2$B$_2$C series \cite{Mueller01}. Actually, the decrease
of T$_c$ roughly scales with the deGennes factor when going
through R = Lu, Tm, Er, Ho and Dy. This has been taken as an
evidence for the presence of a magnetic pair breaking effect that
reduces T$_c$ along the series, following Abrikosov-Gorkov theory
\cite{Mueller01}. However, it does not explain the spectacular
difference between the tunnelling spectra in ErNi$_2$B$_2$C and
TmNi$_2$B$_2$C. They have very different magnetic properties, but
still they are adjacent compounds on the RNi$_2$B$_2$C series,
located closest to the non-magnetic cases. Clearly, a more precise
description, taking into account the peculiarities of each
compound should be very helpful. For instance, the multiband
structure of the Fermi surface
\cite{Dugdale99,Mueller01,Shulga,Martinez03b}, common to the whole
series, has not been theoretically fully taken into account to
explain the behavior of the magnetic compounds. Actually, our
results can be naturally explained if the superconducting gap does
not open on the whole Fermi surface of ErNi$_2$B$_2$C, but only on
a fraction representing roughly half of the overall area. This
would lead to the tunnelling spectra discussed here. It should be
interesting to study also compounds with R = Ho and Dy as they may
show even more extreme behaviors.

It is important to remember that local tunnelling spectroscopy is
a surface sensitive technique. We have measured fourteen high
quality single crystalline samples, prepared in the same manner as
previously measured samples of RNi$_2$B$_2$C by our group (R=Y,
Lu, Tm, \cite{Martinez03b,Suderow01}), in the best possible
conditions, i.e. fresh surfaces broken in-situ in cryogenic vacuum
conditions, and in an experimental set-up with high resolution in
energy that has given radically different results for R = Y, Lu,
Tm. In addition, all past high resolution scanning tunnelling
spectroscopy measurements made in superconductors of coherence
lengths of the order of 10 nm or larger have given results that
are indeed dominated by the bulk properties (see e.g.
\cite{Hess90,Rubio01,Cubitt03,Martinez03b,Suderow01,Rodrigo04,Suderow04}).
So there is no a priori reason to think that the surface may be at
the origin of the observed behavior. But on the other hand,
present knowledge about surface magnetism in compounds with large
exchange field (in comparison to TmNi$_2$B$_2$C) is very poor.
Therefore, there is no way to completely rule out some surface
effect related to magnetism, which would make ErNi$_2$B$_2$C a
different case among the superconductors studied so far with STM.
For instance, if a small ferromagnetic layer nucleates on the
surface below a temperature of the order of T$_c$ or higher, it
could create a sizable local magnetic field, which affects the
superconducting properties at the surface. Another interesting
effect, which creates a high zero bias conductance in the
superconducting tunnelling density of states due to enhanced
magnetic scattering near the surface, has been recently considered
in Refs.\ \cite{Cretinon05,Buzdin05}. This needs to be explored in
more detail and is particularly important in view of planning
phase sensitive experiments in magnetic superconductors
\cite{Tsuei00,Mineev04}, which need typically junctions that
should be very sensitive to surface magnetism.

Summarizing, we have made local scanning tunnelling spectroscopy
measurements on the surface of the rare earth nickel borocarbide
ErNi$_2$B$_2$C, where superconductivity coexists with weak
ferromagnetism, antiferromagnetism and local moment paramagnetism
when increasing temperature up to T$_c$. We directly observe a
finite density of states at the Fermi level as high as half the
normal state value over large areas of the surface.

We acknowledge discussions with A.I. Buzdin, J.P. Brison, F.
Guinea and J. Flouquet. The Laboratorio de Bajas Temperaturas is
associated to the ICMM of the CSIC. Ames Laboratory is operated
for the U. S. Department of Energy by Iowa State University under
Contract No. W-7405-Eng-82. This work was supported by the spanish
MEC (grant FIS-2004-02897), the Comunidad Aut\'onoma de Madrid and
by the Director for Energy Research, Office of Basic Energy
Sciences.


\end{document}